\documentclass[twocolumn,prb,aps,amsfonts]{revtex4-1}
\usepackage{latexsym}
\usepackage{dcolumn}
\usepackage[dvips]{graphicx}
\usepackage{amssymb}
\usepackage{graphicx}
\usepackage{psfrag, color}
\usepackage{verbatim}
\usepackage{amsmath}
\usepackage{epsf}

\begin{document}

%\twocolumn[
%\hsize\textwidth\columnwidth\hsize\csname@twocolumnfalse\endcsname
\draft

\title{Short-range disorder effects on electronic transport in 2D semiconductor structures
  } 
\author{S. Das Sarma$^1$ and E. H. Hwang$^{1,2}$}
\address{$^1$Condensed Matter Theory Center, 
Department of Physics, University of Maryland, College Park,
Maryland  20742-4111 \\
$^2$SKKU Advanced Institute of Nanotechnology and Department of Physics, Sungkyunkwan
University, Suwon, 440-746, Korea} 
\date{\today}

\begin{abstract}
We study theoretically the relative importance of short-range disorder in determining the low-temperature 2D mobility in GaAs-based structures with respect to Coulomb disorder which is known to be the dominant disorder in semiconductor systems. We give results for unscreened and screened short-range disorder effects on 2D mobility in quantum wells and heterostructures, comparing with the results for Coulomb disorder and finding that the asymptotic high-density mobility is always limited by short-range disorder which, in general, becomes effectively stronger with increasing `carrier density' in contrast to Coulomb disorder. We also predict an intriguing re-entrant metal-insulator transition at very high carrier densities in Si-MOSFETs driven by the short-range disorder associated with surface roughness scattering.
\end{abstract}

\maketitle

%\section{introduction}

One of the most spectacular applied physics and electronic materials
advances of the last 40 years has been the 2000-fold enhancement of
the low-temperature carrier mobility in molecular beam epitaxy (MBE)-grown GaAs-based 2D
semiconductor structures. Between 1978, when the modulation doping
technique was introduced by St\"{o}rmer {\it et al.} \cite{stormer} in
2D n-GaAs-AlGaAs heterostructures, and now (i.e., 2014), the
electronic mobility ($\mu$) at low temperatures (below the
Bloch-Gr\"{u}neisen temperature scale, where phonons are unimportant \cite{kawamura})
has increased in 2D GaAs systems from $\mu \approx 2 \times 10^4$
cm$^{2}$/Vs to $\mu \approx 4 \times 10^7$ cm$^{2}$/Vs\; \cite{heiblum}
due to the steady improvement in purity, growth, fabrication, and
processing techniques of individual 2D GaAs samples. As an aside, it
may be pointed out that this increase in low-temperature mobility in
2D GaAs structures is comparable in magnitude to the much better known
Moore's law exponential increase in the room-temperature
microprocessor speed in CMOS
%complementary metalÐoxideÐsemiconductor 
devices achieved over the same time scale
although the qualitative origins of the two enhancements are very
different (better MBE growth and improved modulation doping for
individual 2D n-GaAs structures whereas better packaging and
miniaturization for Si microprocessors).  

In this work, we study theoretically the harmful effects of
short-range disorder scattering in limiting the low-temperature
electronic mobility in 2D GaAs structures, contrasting short-range
disorder scattering with the corresponding long-range Coulomb disorder
scattering arising from quenched charged impurities which has been
extensively studied in the literature \cite{hwang1,hwang2}. It is
well-known that Coulomb disorder arising from unintentional charged
impurities in the background and from the intentional dopants needed
in order to produce  free carriers in semiconductors is the main
mobility-limiting scattering mechanism in high-mobility 2D GaAs
structures. However, at higher carrier densities, where Coulomb
disorder is strongly screened out, other weaker and short-ranged
scattering mechanisms arising from alloy disorder, interface
roughness, neutral impurities, and defects could play a quantitative
role in determining the 2D mobility in high-mobility 2D GaAs
systems. Theoretically studying the quantitative effect of such
short-range disorder scattering is the goal of this work. 

Our theory uses the $T=0$ Boltzmann transport equation for calculating
the carrier mobility ($\mu$) limited by individual scattering
mechanisms by directly calculating the relevant transport relaxation
or scattering time ($\tau$) with the mobility being defined as $\mu =
e \tau/m$ and the conductivity $\sigma$ given by $\sigma = n e \mu$,
where $n$ is the 2D carrier density (and $m$ the carrier effective
mass). Within the leading-order Born approximation the disorder
induced scattering rate $\tau^{-1}$ is given (at $T=0$) by \cite{andormp,dassarma2011}
\begin{eqnarray}
\frac{1}{\tau_i(k)} = \frac{2\pi}{\hbar} \int \frac{d^2k'}{(2\pi)^2} & &  \int_{-\infty}^{\infty} dz N_i(z) \frac{ \left |V_i(q,z) \right |^2}{\epsilon(q)^2} \nonumber  \\
& &\times   (1-\cos\theta) \delta(E_k-E_{k'}).
\end{eqnarray}
Here `$i$' denotes the $i$-th type of disorder characterized by a
random distribution of scattering centers with density $N_i(z)$ along
the $z$-direction nomal to the 2D layer with $V_i$ being the bare
impurity potential and $\epsilon(q)$ the static RPA dielectric
screening function for the 2D carrier system. If screening is
irrelevant as it is for most short-range disorder, we take
$\epsilon(q) =1$. Also, $E_k = \hbar^2k^2/2m$, is the 2D carrier
energy. The functions $V_i$, $\epsilon$, $etc.$ are to be calculated
in the appropriate basis of the quasi-2D quantized confinement
wavefunction (along the $z$ direction) of the 2D semiconductor
structure so that they are really matrix elements in the appropriate
ground subband of the 2D system. For Coulomb disorder, $V_i(q,z) =
(2\pi e^2Z_i/\kappa q) e^{-qz}$, where $\kappa$ is the background
lattice dielectric constant and $Z_i$ is the impurity charge strength
of the charged center producing the disorder. For short-range
disorder, we assume with no loss of generality, $V_i \equiv V_0
\delta({\bf r} -{\bf r}_i)\delta(z-z_i)$, an uncorrelated delta-function white noise
disorder characterized by a potential strength $V_0$ (as well as an
impurity density $N_i$). 
It is then straightforward to calculate the short-range scattering rate
\begin{equation}
\frac{1}{\tau(k)} = mN_iV_0^2C \int_0^2 \frac{dx}{\sqrt{4-x^2}}\frac{x^2}{\epsilon(kx)^2},
\end{equation}
wgere $C= \int |\xi_0(z)|^4 dz$ with $\xi_0(z)$ being the ground state
quasi-2D confining wavefunction defining the quantum bound state of
the 2D system along the $z$-direction, i.e., the total single-particle
ground-state wavefunction is given by $\psi_0({\bf r},z) = \xi_0(z)
e^{i{\bf k} \cdot {\bf r}}$, where {\bf r}, {\bf k} are 2D position and wave vector
respectively. 
 
%%%%%%%%%%%%%%%%%%%%%

If screening is neglected and the 2D system is approximated to be an
idealized zero-thickness layer, then $\epsilon=1$ and $C=1$, so that
the scattering rate $[\tau(k)]^{-1} = m N_i V_0^2$ is a constant
determined only by the strength of the short-range disorder (and
carrier effective mass), becoming completely independent of 2D wave
vector and carrier density. (In practice, there might be some carrier
density dependence even in the absence of screening because of the
quasi-2D form-factor associated with the confinement wavefunction.) If
it is appropriate to screen the short-range disorder using the static
RPA dielectric function $\epsilon(q) = 1+ q_{TF}/q$, where $q_{TF}$ is
the 2D screening wave vector, Eq.~(2) gives 
\begin{equation}
\frac{1}{\tau(k)} = m N_iV_0^2 C k^2 \int_0^2 \frac{dx}{\sqrt{4-x^2}} \frac{x^4}{(kx + q_{TF})^2}.
\end{equation}
Putting $k=k_F \propto \sqrt{n}$ as appropriate at $T=0$, Eq.~(3)
implies that the 2D mobility $\mu$ [$\propto \tau(k_F)$] limited by
screened short-range disorder has the following high and low-density
behavior (ignoring any density dependence arising from the confinement
wavefucntion effects, i.e., in the strict 2D limit) 
\begin{eqnarray}
\mu & \sim & n^0, \;\;\;\;\;\;\; q_{TF} \ll 2k_F \nonumber \\
        & \sim & n^{-1}, \;\;\;\;\; q_{TF} \gg 2k_F.
\end{eqnarray}        
Thus, the high-density limit of the screened situation is the same as
the unscreened case of no density dependence (since $q_{TF} \ll 2 k_F$
in the high-density limit), but the low-density (and strongly screened
$q_{TF} \gg 2k_F$) situation produces the somewhat counter-intuitive
$\mu \sim n^{-1}$ dependence where increasing (decreasing) 2D carrier
density suppresses (enhances) the 2D mobility constrained by
short-range disorder scattering. By contrast, the Coulomb disorder
limited mobility behaves as $\mu \sim n^{\alpha(n)}$ with
$\alpha(n\rightarrow 0) \sim 0$ and $\alpha(n\rightarrow \infty) \sim
1$ or 3/2 depending respectively on whether the charged impurities are near or far from the 2D
layer \cite{hwang2}. Thus, Coulomb disorder limited 2D mobility always
increases with increasing 2D carrier density whereas the short-range
disorder limited 2D mobility is either a constant independent of
density or decreases with increasing density. This immediately implies
that short-range disorder becomes increasingly more important at
higher carrier densities where Coulomb disorder is increasingly
suppressed. (In fact, similar transport behavior occurs also in graphene \cite{dassarma2011} where the low- to intermediate-density mobility is a constant because of Coulomb disorder whereas the high-density mobility decreases with increasing carrier density because of short-range scattering.)
Thus, the low-temperature high-density maximum mobility of
all 2D systems is eventually always limited by short-range scattering
since the Coulomb disorder limited scattering is strongly and
monotonically suppressed with increasing density. This behavior is
well-known in 2D Si-MOSFETs \cite{andormp} and in graphene
\cite{hwanggraphene,columbia}, but is true for 2D GaAs systems also
where the mobility must saturate (or even decrease) at high enough
carrier density because any existing short-range disorder (no matter
how small) arising from alloy disorder or interface roughness or
neutral defects would eventually dominate the high-density
mobility once the Coulomb scattering effects have become negligible. (The 2D Coulomb disorder case has recently been discussed in
depth by us elsewhere \cite{hwang1,hwang2}.) 

We note that the fact that the low (high)-density regime is more strongly (weakly) screened than the high (low)-density regime follows simply from the Fermi wave vector $k_F$ going as $n^{1/2}$ in 2D versus the screening wave vector $q_{TF}$ going as a constant -- the same situation applies in 3D systems too where $k_F \sim n^{1/3}$ and $q_{TF} \sim n^{1/6}$.  This is simply a universal feature of the dimensionless quantum screening parameter $q_{TF}/2k_F$ increasing with decreasing carrier density in a quantum degenerate electron gas -- obviously, the absolute magnitude of screening (i.e. $q_{TF}$ itself) always increases with increasing density both in quantum and classical situations.  While any long-range Coulomb disorder (arising from random charged impurity scattering) must always be screened, the short-range disorder may or may not always be screened depending on its physical origin -- for example, the short-range disorder arising from  interface roughness (atomic defects or vacancies) should typically be screened (unscreened).

We now present our detailed numerical results for 2D mobility limited
by short-range scattering comparing and contrasting with the
corresponding Coulomb disorder situation. Our numerical results (all
at $T=0$) include full effects of the appropriate quais-2D confinement
wavefunction in our calculations, both for GaAs quantum wells [where
the well-width `$a$' determines $\xi_0(z)$] and heterostructures
[where the ground-state confinement wavefunction $\xi_0(z)$ depends
explicitly on the carrier density $n$ through the self-consistency
effect \cite{stern,andormp}, thus providing an additional density
dependence of the mobility]. 
Our use of Born approximation in the Boltzmann transport theory is justified by the high carrier density regime of our interest, and is also borne out by the extensive success of the Born approximation in calculating 2D semiconductor transport properties over the last 40 years, both for long-range Coulomb disorder and  short-range interface roughness disorder. \cite{andormp}  We mention also that the parameters used for our transport calculations are realistic numbers for Si MOS structures \cite{andormp}, but not for GaAs-based 2D systems where the interface roughness scattering is much weaker than in Si MOSFETs since the GaAs-AlGaAs interface has very little roughness with GaAs and AlGaAs being almost lattice-matched.  The other differences between Si and GaAs systems are the different effective masses, background dielectric constants, and lower impurity disorder in GaAs.

%We mention that the total mobility of the system obeys Matthiessen's rule precisely at $T=0$K (only approximately for $T\neq 0$), and is thus given by $\mu^{-1} = \sum_i \mu_i^{-1}$ where $\mu_i$ denotes the mobility due to the $i$-th independent scattering mechanism, implying that when a specific scattering mechanism dominates (e.g., Coulomb and short-range disorder at lower and higher carrier densities, respectively) the other mechanisms become irrelevant. 

In Figs.~1--3 we show respectively our numerical results for the
mobility ($\mu$) as a function of carrier density ($n$) in GaAs-AlGaAs
heterostructures (Fig. 1) and GaAs-AlGaAs quantum wells (Figs. 2 and
3). In each case, $\mu(n)$ as well as the density scaling exponent
$\alpha(n) \equiv d \ln \mu/ d \ln n$ of mobility\cite{hwang2}, i.e., $\mu \sim
n^{\alpha}$, are both shown for several different models of long- and
short-range disorder scattering: unscreened and screened Coulomb
disorder; unscreened and screened neutral short-range disorder. In
Figs.~1 (heterostructure) and 2 (quantum well), the impurities are
located inside the 2D electron layer whereas in Fig.~3 (quantum well)
the impurities are located at the interface. The absoulte magnitude of
the mobilities are not particularly relevant since the impurity
densities (or the scattering strength) are not known in general, but
the relative variations in the mobility as a function of density for
various scattering mechanisms provide a meaningful comparison. We
adjusted the various unknown impurity parameters (e.g. $N_i$ for
Coulomb disorder and $N_i V_0^2$ for short-ranged disorder) so that
all the results for different scattering mechanisms can be fitted
within the same figure. It is easy to obtain the mobility limited by
arbitrary scattering strength (i.e. different values of $N_i$ and/or
$V_0^2$) simply by appropriately scaling our results shown in
Figs.~1--3, remembering that $\mu^{-1} \propto N_i$ (or $N_iV_0^2$). 

% All Figure files are in the Bell_Lab directory.
%%%%%%%%%%%%%%%% Fig. 1 %%%%%%%%%%%%%%%%%%%%%%%%%%%%
\begin{figure}[t]
	\centering
	\includegraphics[width=1.\columnwidth]{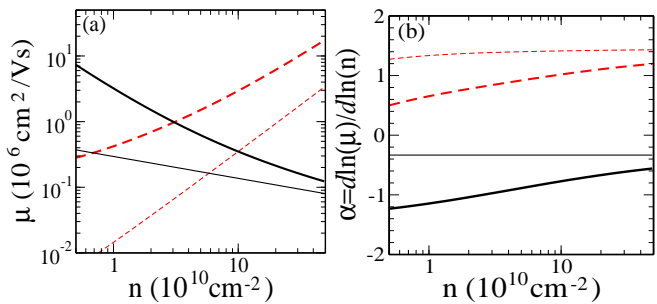}
	\caption{Calculated (a) mobility and (b) exponent as a function of density for GaAs-AlGaAs heterostructures. The solid (dashed) lines show the results calculated with short-range (long-range) disorder scattering. The thick (thin) lines  represents the results for screened (unscreened) disorder. All impurities are located inside the 2D electron layer.
}
\end{figure}

%%%%%%%%%%%%%%%% Fig. 2 %%%%%%%%%%%%%%%%%%%%%%%%%%%%
\begin{figure}[b]
	\centering
	\includegraphics[width=1.\columnwidth]{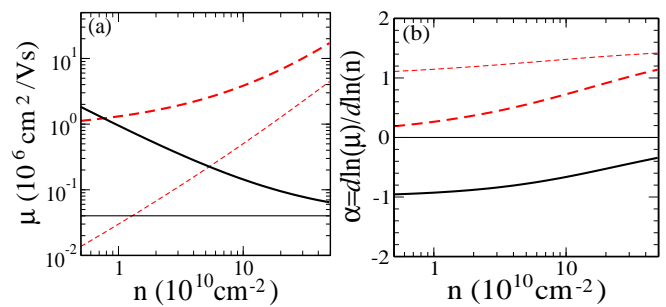}
	\caption{Calculated (a) mobility and (b) exponent as a
          function of density for GaAs-AlGaAs quantum wells with well
          width $a=200$\AA. The solid (dashed) lines show the results
          calculated with short-range (long-range) disorder
          scattering. The thick (thin) lines  represents the results
          for screened (unscreened) disorder. All impurities are
          located inside the 2D quantum well. 
}
\end{figure}

The most important qualitative results of Figs.~1--3 are the
following: (i) The mobility often, but not always, could decrease with
increasing density for short-range disorder scattering; (ii) the
high-density mobility is always dominated by short-range disorder;
(iii) a consequence of the last statement is that at sufficiently high
density $\alpha(n)$, where $\mu(n) \sim n^{\alpha}$, will either
become zero or become negative; (iv) the detailed quantitative aspects
of short-range disorder effect on the mobility would depend on many
(unknown) microscopic details such as the location of the disorder (at
the interface or in the layer), screened or not, quantum well or
heterostructure;  (v) the 2D mobility, when limited by short-range
disorder in the channel, has an exponent $-2 \alt  
\alpha \alt 0$ which is qualitatively different from the exponent
$\alpha$ limited by Coulomb disorder ($\alpha > 0.5$) -- whether
$\alpha$ is actually zero or negative depends on whether the
short-range disorder is screened or not; (vi) the mobility decreases
linearly with $N_i$, and thus it decreases linearly with the number of
Al atoms (i.e. Al density) in the channel assuming that the mobility
is limited by alloy disorder scattering \cite{manfra}.

%%%%%%%%%%%%%%%% Fig. 3 %%%%%%%%%%%%%%%%%%%%%%%%%%%%
\begin{figure}[t]
	\centering
	\includegraphics[width=1.\columnwidth]{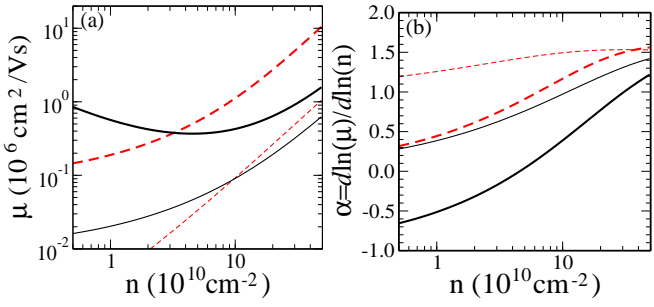}
	\caption{The same as Fig. 2, but the impurities are located at the interface.
}
\end{figure}

We note that it is straightforward to derive the following dependence
of mobility limited by unscreened short-range disorder on the quasi-2D
form-factor arising from the wavefunction confinement effect; 
\begin{equation}
\frac{1}{\tau_k} = mN_iV_0^2 \left \{
\begin{matrix}
3b/16, & {\rm \;\;\;  heterostructure}  \cr
3/2a, & {\rm  quantum \; well} \cr
\end{matrix} 
\right .
\end{equation}
where `$a$' is the quantum well width and `$b$' is the variational
parameters defining $\xi_0(z)$ ($\sim z e^{-bz/2}$) which is given by
$b \propto n^{1/3}$ in the simple Stern-Howard variational
approximation \cite{andormp}. In general, $b \propto n^{1/3}$ and $a \propto n^0$,
and thus $\mu \sim n^{-1/3}$ (heterostructure) and $\mu \sim n^{0}$
(quantum well) for the unscreened short-range disorder. In the
presence of screening, an extra factor of $1/k_F^2$ comes in (see
Eq.~3) so that we get $\mu \sim n^{-4/3}$ (heterostructure) and $\mu
\sim n^{-1}$ (quantum well) in the $q_{TF} \gg 2k_F$ limit of screened
short-range disorder. For $q_{TF} \ll 2k_F$, the screening effect
disappears as discussed earlier. 

Finally, we conclude by making an intriguing prediction about a
re-entrant metal-insulator transition which should occur in 2D
Si-MOSFETs at a very high carrier densities ($ > 10^{13}$ cm$^{-2}$)
driven entirely by the short-range disorder scattering associated with
the surface roughness at the Si-SiO$_2$ interface which is known to be
the dominant high-density mobility-limiting mechanism \cite{andormp}.  
As shown in Fig.~4, the Si MOSFET mobility $\mu(n)$ at first increases
(for $n < n_m$) with increasing carrier density reaching a maximum
sample-dependent value $\mu_m$ at some characteristic sample-dependent
density $n_m$ with $\mu(n>n_m)$ decreasing with increasing carrier
density at high density\cite{andormp}. The numerical results shown in Fig.~4 are
easily understood based on the realistic model of just two scattering
mechanisms, leading to mobilities limited by Coulomb impurities
($\mu_{CI}$) and surface roughness ($\mu_{SR}$) dominating at low ($n
< n_m$) and high ($n > n_m$) densities, respectively 
\begin{equation}
\mu = \left [ \mu_{CI}^{-1}(n) + \mu_{SR}^{-1}(n) \right ]^{-1},
\end{equation}
with $\mu_{CI}(n) \sim n^{\alpha_{CI}(n)}$, where $\alpha_{CI} \sim 0.4$, and $\mu_{SR}(n) \sim n^{\alpha_{SR}(n)}$, where $\alpha_{SR} \sim -2$. From Eq. (6), we then get 
\begin{equation}
\mu=(An^{-0.4} + Bn^2)^{-1},
\end{equation}
where $A$ and $B$ are constants (i.e. roughly independent of carrier
density) which depend on the details of Coulomb disorder and surface
roughness disorder, respectively. Equation (7) implies $\mu(n
\rightarrow 0) \sim n^{0.4}$ and $\mu(n \rightarrow \infty) \sim
n^{-2}$, and thus $\mu(n)$ has a maximum at a disorder-dependent
non-universal density $n_m(A,B)$ with the value $\mu_m = \mu(n_m)$. It
is easy to show $n_m \approx 1.7 (A/B)^{0.4} \propto 1.7
(N_{CI}/V_{SR}^2)^{0.4}$ where $N_{CI}$ and $V_{SR}^2$ are the charged
impurity disorder strength and surface roughness scattering strength,
respectively. Thus, $n_m$ $(\mu_m)$ increases (decreases) with increasing
charged impurity density. All of these behaviors are clearly manifest
in our Fig.~4 where we show realistic numerical results for two
situations with low and high values of $N_{CI}$ (thus corresponding to
a high- and a low-quality MOSFET sample, respectively) using the same
surface roughness disorder parameters. 
The two sets of results in Fig. 4 [i.e.,  4(a) and 4(b)] correspond to using two different sets of surface-roughness materials parameters at the Si-SiO$_2$ interface which are in general not independently known and often inferred based on the modeling of transport measurements.

%%%%%%%%%%%%%%%% Fig. 3 %%%%%%%%%%%%%%%%%%%%%%%%%%%%
\begin{figure}[t]
	\centering
	\includegraphics[width=1.\columnwidth]{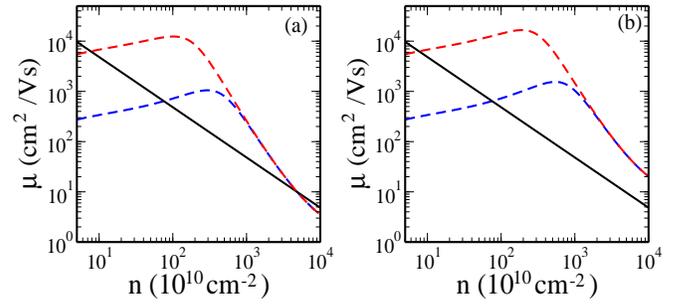}
	\caption{Calculated mobility in Si-MOSFETs as a function of density. The solid line represents $\mu=(2e/h)(1/n)$, which is equivalent to $k_Fl=1$. The top (bottom) dashed line (in each figure) represents the mobility of a high (low) quality Si-MOSFET (i.e., with large or small amount of Coulomb disorder).  The surface roughness parameters, $\Delta=10$ \AA \; and 4.5 \AA,
are used in (a) and (b), respectively (see ref. [\onlinecite{andormp}] for details).
The figures 4(a) and (b) correspond respectively to roughness parameters allowing (or not) a re-entrant transition to a high-density insulating phase -- see the text for details.	
	}
\end{figure}

What we summarize above is known in some form in the literature on
Si-MOSFET transport properties\cite{andormp}, but now we make a
striking new observation which seems to have been completely
overlooked in the extensive literature on MOSFETs. We note that the
transport mean free path ($l$), $l=v_F \tau = v_F m \mu/e$, with $v_F
= p_F/m =\hbar k_F/m$ being the Fermi velocity, is very small in
Si-MOSFETs both at low carrier density (where Coulomb scattering
dominates) and at high carrier density (where short-range surface
roughness disorder dominates). It is well-known that Si-MOSFETs
universally undergo a low-density metal-insulator transition (MIT) at
a critical density $n_{c1}$ ($<n_m$), dominated by charged impurity
scattering, where the mean free path becomes short enough so that the
Ioffe-Regel-Mott criterion $k_F l = 1$ is satisfied. This low-density
MIT in Si-MOSFETs has been studied extensively in the literature over
the last forty years \cite{andormp,abrahams,dassarma2011}. What we
predict here is that the 
Ioffe-Regel-Mott criterion may also be satisfied at high density ($n >
n_m$) where very strong short-range surface roughness scattering would
eventually reduce the mobility to a low enough value ($< 100$
cm$^2$/Vs) so that $k_F l = 1$ would be satisfied at a second critical
density $n_{c2}$ ($> n_{c1}$) with the insulating state being
re-entrant for $n>n_{c2}$ (as well as for $n<n_{c1}$) with the
intermediate $n_{c1} <n <n_{c2}$ density regime being metallic. We
show the $k_Fl = 1$, which is equivalent to $\mu = (2e/h)(1/n)$
line, in our Fig. 4 to emphasize the fact that there can, in
principle, be two solutions for the integral equation $k_F l = 1$ at
two densities, depending on the details of disorder. Our extensive
numerical investigations with many possible realistic Si-MOSFET
disorder parameters we find that $n_{c2} \agt 10^{13}-10^{14}$
cm$^{-2}$ in general with $n_{c2} \approx 3-5 \times 10^{13}$
cm$^{-2}$ being the most likely value for most realistic Si-MOSFET
samples. By contrast $n_{c1} \agt 10^{11}-10^{12}$ cm$^{-2}$ typically
depending on the amount of Coulomb disorder in the system. 
Of course, it is also possible that the surface roughness scattering is never strong enough to induce $k_Fl=1$ condition at any carrier density in which case there will be no re-entrant transition to an insulating phase at high carrier density as shown in Fig. 4(b).

If low-temperature transport measurements are feasible in Si-MOSFETs at
carrier densities above $10^{13}$ cm$^{-2}$, we believe that our
predicted re-entrant metal-insulator transition could be observed
experimentally. We emphasize, however, that the integral equation $\mu(n)
= (2e/h)(1/n)$ may have zero, one [Fig. 4(b)], or two [Fig. 4(a)] solutions in general
depending on the details and the relative strengths of the applicable
long- and short-range disorder in the system, given that $\mu(n) \sim
n^{-2}$ for $n \rightarrow \infty$ and $\sim n^{0.4}$ for $n
\rightarrow 0$. The three solutions correspond respectively to the
system being always localized because of very large disorder ($k_Fl <
1$ for all density), having an insulating phase at low density $n<n_{c1}$
driven by Coulomb disorder and a metallic phase for all $n>n_{c1}$ [as in Fig. 4(b)], and
the re-entrant insulating case with a second insulator for $n>n_{c2}$ [as in Fig. 4(a)]
driven by strong surface roughness disorder. The first two situations
are well-known in the MOSFET literature, but the re-entrant
high-density insulating phase predicted here has not yet been reported
experimentally to the best of our knowledge. 

This work is supported by LPS-CMTC, Microsoft Q, and  Basic Science Research Program through the National Research Foundation of Korea Grant funded by the Ministry of Science, ICT \& Future Planning (2009-0083540).

\end{document}